\documentclass[
 reprint,
 superscriptaddress,
 amsmath,amssymb,
 aps,
pra,
showkeys
]{revtex4-2}

\usepackage{soul}
\usepackage{float}
\usepackage{multibib}
\usepackage{hyperref}
\usepackage{xcolor}
\hypersetup{
    colorlinks,
    linkcolor={blue!50!blue},
    citecolor={blue!50!blue},
    urlcolor={blue!80!black}
}
\usepackage{siunitx}
\usepackage{import}
\usepackage{placeins}
\usepackage{xcolor}
\usepackage{lmodern}
\usepackage[braket, qm]{qcircuit}
\usepackage{graphicx}
\usepackage{dcolumn}
\usepackage{bm}
\usepackage{comment}
\usepackage[english]{babel}
\usepackage{blindtext}
\usepackage{physics}  
\usepackage{multirow}
\usepackage{array}
\usepackage{enumerate}

\usepackage{booktabs}
\usepackage{ulem}
\usepackage{enumitem}

\begin{document}
\title{Advancing the hBN Defects Database through Photophysical Characterization of Bulk hBN}  

\author{Chanaprom Cholsuk}
\email{chanaprom.cholsuk@tum.de}
\affiliation{Department of Computer Engineering, TUM School of Computation, Information and Technology, Technical University of Munich, 80333 Munich, Germany}
\affiliation{Munich Center for Quantum Science and Technology (MCQST), 80799 Munich, Germany}

\author{Sujin Suwanna}
\affiliation{Optical and Quantum Physics Laboratory, Department of Physics, Faculty of Science, Mahidol University, Bangkok 10400, Thailand}

\author{Tobias Vogl}%
\email{tobias.vogl@tum.de}
\affiliation{Department of Computer Engineering, TUM School of Computation, Information and Technology, Technical University of Munich, 80333 Munich, Germany}
\affiliation{Munich Center for Quantum Science and Technology (MCQST), 80799 Munich, Germany}
\affiliation{Abbe Center of Photonics, Institute of Applied Physics, Friedrich Schiller University Jena, 07745 Jena, Germany}

\date{\today}

\begin{abstract}
Quantum emitters in hexagonal boron nitride (hBN) have gained significant attention due to a wide range of defects that offer high quantum efficiency and single-photon purity at room temperature. Most theoretical studies on hBN defects simulate monolayers, as this is computationally cheaper than calculating bulk structures. However, most experimental studies are carried out on multilayer to bulk hBN, which creates additional possibilities for discrepancies between theory and experiment. In this work, we present an extended database of hBN defects that includes a comprehensive set of bulk hBN defects along with their excited-state photophysical properties. The database features over 120 neutral defects, systematically evaluated across charge states ranging from –2 to +2 (600 defects in total). For each defect, the most stable charge and spin configurations are identified and used to compute the zero-phonon line, photoluminescence spectrum, absorption spectrum, Huang-Rhys (HR) factor, interactive radiative lifetimes, transition dipole moments, and polarization characteristics. Our analysis reveals that the electron–phonon coupling strength is primarily influenced by the presence of vacancies, which tend to induce stronger lattice distortions and broaden phonon sidebands. Additionally, correlation analysis shows that while most properties are independent, the HR factor strongly correlates with the configuration coordinates. All data are publicly available at \url{https://h-bn.info}, along with a new application programming interface (API) to facilitate integration with machine learning workflows. This database is therefore designed to bridge the gap between theory and experiment, aid in the reliable identification of quantum emitters, and support the development of machine-learning-driven approaches in quantum materials research.
\end{abstract}

\keywords{hexagonal boron nitride, defects, density functional theory, database, bulk}

\maketitle

\section{Introduction}
Solid‐state quantum emitters have become essential building blocks for optical quantum technologies, owing to their ability to generate on‐demand single photons that serve as flying qubits in applications such as quantum communication \cite{10.1103/RevModPhys.74.145,10.1002/qute.202300343}, quantum memory \cite{10.1103/PhysRevX.6.021040,10.1002/adom.202302760,10.1063/5.0188597}, quantum computing \cite{10.1038/nature08812,10.1063/5.0007444}, and quantum sensing \cite{10.1038/s41563-024-01887-z,10.1038/s41467-021-24725-1}. A variety of platforms, including quantum dots \cite{10.1038/nnano.2017.218,10.1515/nanoph-2019-0007}, nitrogen‐vacancy (NV) centers in diamonds \cite{10.1016/j.physrep.2013.02.001,10.1126/science.1139831}, silicon carbide (SiC) \cite{10.1364/OPTICA.3.000768}, two‐dimensional transition‐metal dichalcogenides (2D TMDs) \cite{10.1038/s41467-019-10632-z,10.3390/nano13091501}, and defects in hexagonal boron nitride (hBN) \cite{10.1038/nnano.2015.242,10.1364/OPTICA.6.001084}, have demonstrated single-photon emission under different operating conditions. While epitaxial quantum dots exhibit excellent photon purity, they typically require cryogenic operation \cite{10.1038/nnano.2017.218,10.1515/nanoph-2019-0007}. Conversely, bulk crystal defects can function at room temperature but suffer from limited photon out-coupling efficiency due to their three‐dimensional geometry \cite{10.1039/c9nr04269e}. Two-dimensional materials thus offer a compelling compromise.~That is, their atomically thin form reduces the total internal reflection; hence, enhancing the out-coupling efficiency.\\
\begin{figure*}
    \centering
    \includegraphics[width=1\linewidth]{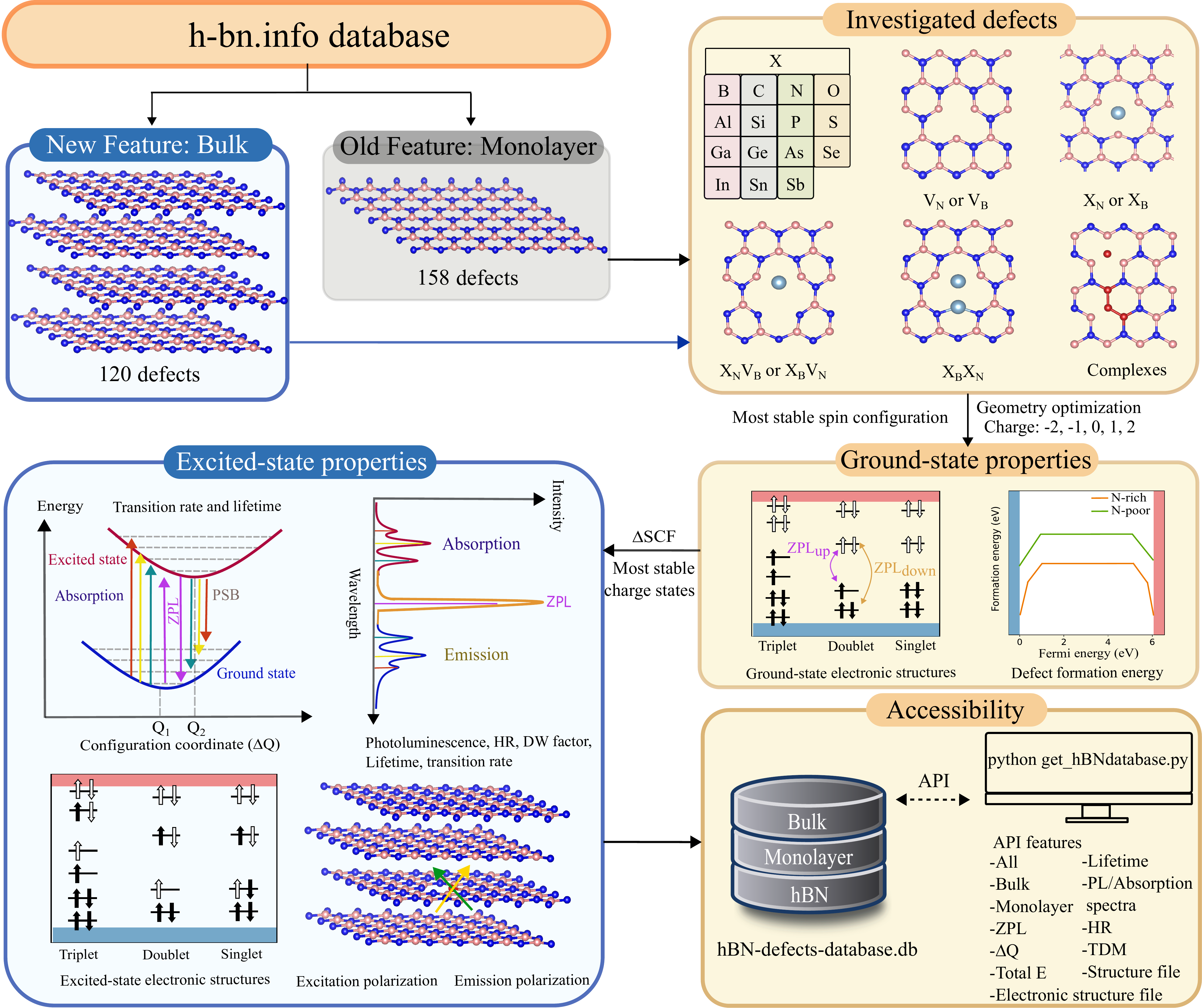}
    \caption{Overview of the \url{h-bn.info} database workflow. The updated database extends our previous work on monolayer hBN by introducing a new feature: bulk hBN defect structures. Investigated defects include native vacancies, substitutional impurities (from groups III–VI), and various complexes, with charge states ranging from –2 to +2. For each defect, the most stable charge and spin configurations are identified. Ground-state properties such as defect formation energies and electronic structures are computed, followed by excited-state properties including~ZPL, PL and absorption spectra, configuration coordinates, HR and DW factors, radiative lifetimes, and transition dipole polarizations. All data are compiled into a structured database with an accessible API supporting Python-based queries for integration with ML and data-driven research workflows. The color scheme is defined as follows: blue blocks represent features available only for bulk, while yellow blocks represent features available for both bulk and monolayer.}
    \label{fig:database_diagram}
\end{figure*}
\indent Among 2D materials, TMD emitters have shown promise, but to date, most demonstrations remain confined to temperatures below 15 K \cite{10.1038/s41467-019-10632-z,10.3390/nano13091501}. In contrast, color centers in hBN operate reliably at room temperature, exhibit high quantum efficiency \cite{10.1364/OPTICA.6.001084}, robustness against aging \cite{10.1021/acsphotonics.8b00127,10.1016/j.physe.2020.114251}, stability \cite{10.1021/acsphotonics.7b00086}, and benefit from high out-coupling due to the atomically thin host lattice \cite{10.1038/nnano.2015.242}. In hBN, a single photon is emitted when an optically excited electron relaxes through a two‐level defect state localized within the wide band gap ($\sim$6 eV) of the material. Given the large band gap, hBN has been proven to accommodate a multitude of defect‐induced mid‐gap states, giving rise to a rich spectrum of optical transitions \cite{10.3390/nano12142427}. Experimentally, quantum emitters in hBN have been observed at photon energies ranging near 1 eV \cite{10.1063/5.0008242}, 2 eV \cite{10.1103/PhysRevB.98.081414,arxiv:2507.06783}, 3 eV \cite{10.1021/acsphotonics.2c00631}, and 4 eV \cite{10.1063/1.5124153}. This wide range of emission energies makes defect identification a central challenge in hBN research. Currently, only the V$_\text{B}^{-1}$ center has a well-established match between theory and experiment \cite{10.1038/s41563-020-0619-6}. For other proposed emitters such as C$_\text{B}$V$_\text{N}$ \cite{10.1088/2053-1583/ab8f61,10.3389/frqst.2022.1007756}, C$_\text{B}$C$_\text{N}$ \cite{10.1063/1.5124153}, C$_\text{B}$C$_\text{N}$C$_\text{N}$ \cite{10.1103/PhysRevMaterials.6.L042201}, C$_\text{B}$C$_\text{N}$C$_\text{B}$C$_\text{N}$ \cite{10.1063/5.0147560,10.1021/acsnano.3c08940,10.1021/acs.jpclett.3c01475,10.1021/acsanm.4c02722}, O$_\text{N}$V$_\text{B}^{-1}$ \cite{10.1021/acs.jpclett.2c02687}, a clear consensus of all photophysical properties between experiment and theory remains elusive.\\
\indent This challenge is based on the fact that many different defects can produce similar signatures, such as overlapping zero-phonon line (ZPL) energies, comparable excited-state lifetimes, similar EPR spectra, and Raman features \cite{arxiv:2502.21118}. A major bottleneck in resolving this issue is the absence of a comprehensive and systematic database covering the photophysical properties of a large set of hBN defects. Therefore, robust defect identification requires a broad and detailed characterization of a wide range of properties across many possible defects \cite{10.1021/acs.jpclett.3c01475}. To mitigate the high computational demand associated with first-principles calculations, several machine learning (ML) approaches have been developed to predict photophysical properties across large defect spaces \cite{10.1021/acsnano.0c05267,10.1021/acs.jpcc.0c05831}. However, these models still rely on DFT benchmarks for validation, as the photophysical properties of hBN defects do not follow simple patterns based on chemical type, symmetry, or defect class. The inherent variability and randomness in defect-related optical properties thus necessitate a comprehensive, DFT-validated repository that includes both ground- and excited-state characteristics to support reliable identification and interpretation of experimental results.\\
\indent To address this gap, we previously introduced a hBN defect database \cite{10.1021/acs.jpcc.4c03404}. However, it was limited to only monolayer hBN and focused primarily on ground-state electronic properties. Since most experimental studies are conducted on bulk hBN samples, there is a pressing need to extend the theoretical framework to suitably accommodate experimental conditions. In this work, we therefore expand the database to include defect properties in bulk hBN and incorporate excited-state photophysical properties. In contrast to our earlier version, as summarized in Fig.~\ref{fig:database_diagram}, the newly developed database includes new defects and an additional analysis of charge stability, where defect charge states ranging from $-$2 to $+$2 are systematically evaluated. For each defect, the most stable charge and spin configurations are identified and subsequently used for detailed calculations of both ground-state and excited-state properties. All computed data are compiled into a publicly accessible database available at \url{https://h-bn.info}. Importantly, the database now provides comprehensive data for both bulk and monolayer hBN, allowing users to select the most relevant system for their research needs. As machine learning models are becoming increasingly popular in materials research, we also provide an application programming interface (API) for automated data retrieval to support broader usability and facilitate integration with ML workflows. As a result, this work serves as a vital resource for guiding experimental findings, supporting defect identification, and facilitating data-driven approaches in quantum materials research.

\section{Method}
This section describes the methodology used to generate the data presented in this work. While each method has been described already in Ref.~\cite{10.1021/acs.jpcc.4c03404}, we still review the details here for the sake of completeness. Each block in Fig.~\ref{fig:database_diagram} corresponds to a step in the computational pipeline and is detailed below.

\subsection{DFT calculation details}
All first-principles calculations in this work were performed using the Vienna Ab initio Simulation Package (VASP) \cite{vasp1,vasp2}, employing the projector augmented-wave (PAW) method for treating core–valence interactions \cite{paw,paw2}. To accurately capture the electronic structure of wide-bandgap hBN, we employed the screened hybrid functional HSE06, modified with a Hartree–Fock exchange mixing parameter $\alpha$=0.32, which yields a calculated band gap of 6.09 eV for hBN, in excellent agreement with experiment \cite{10.1038/s41467-019-10610-5}. A single $\Gamma$-point was used for Brillouin-zone sampling in all defect calculations, and a plane-wave cutoff energy of 500 eV was applied. Spin polarization was included throughout to ensure spin conservation in electronic transitions.\\
\indent Experimentally, bulk hBN can adopt several stacking sequences, most notably AA' (boron atoms in one layer sit directly above nitrogen atoms in the adjacent layer) and AB (each successive layer is translated by one in‐plane bond length) \cite{10.1088/2053-1583/ab0e24}. In this study, we employ the AA' configuration exclusively, as total-energy calculations based on hybrid functional identify it to be slightly more stable than the AB stacking \cite{10.1021/acs.jpcc.8b09087,10.1103/PhysRevLett.111.036104}. The van der Waals interactions were included via the D3 correction of Grimme et al.~\cite{10.1063/1.3382344}.

\subsection{Selection of investigated defects}
To systematically explore defects in bulk hBN, we considered native defects, X$_\text{N}$, X$_\text{B}$, X$_\text{N}$V$_\text{B}$, X$_\text{B}$V$_\text{N}$, X$_\text{B}$X$_\text{N}$, and complexes where X is an element from groups III - VI indicated in Fig.~\ref{fig:database_diagram}.

All defect structures were generated within a 6$\times$6$\times$2 supercell (288 atoms), as illustrated in Fig.~\ref{fig:database_diagram}, to minimize spurious defect–defect interactions. Ionic positions were relaxed until residual forces fell below 10$^{-2}$ eV/\AA~ and total energy changes were smaller than 10$^{-4}$ eV. Note that full relaxation of both atomic positions and lattice parameters was also tested and found to cause zero to 20 nm variation in the ZPL, depending on the defect type. Given this minor effect, we chose to relax only the ionic positions. In total, 124 distinct neutral defect configurations were constructed and carried forward for subsequent charge stability analysis.

\subsection{Ground-state properties}
\subsubsection{Ground-state electronic structures}
For hBN defects, the most commonly encountered spin configurations include singlet, doublet, and triplet states, each corresponding to a different number of unpaired electrons. In this work, we determine the preferred spin configuration for each defect by comparing the total energies of all relevant spin states. Only the lowest-energy configuration is selected for subsequent calculations.

\subsubsection{Defect formation energy}
Defect formation energy is a key quantity used to determine the most stable charge state of a defect under given thermodynamic conditions. For each defect in this study, charge states ranging from $-$2 to $+$2 were systematically evaluated by computing their formation energies using the following expression
\begin{eqnarray}
    E^{f}[X^{q}] &=& E_{\text{tot}}[X^{q}] - E_{\text{tot}}[\text{hBN}] - \sum_{i}n_i \mu_i \nonumber \\
    &+& q(\epsilon_{\text{vbm}} + \epsilon_{\text{Fermi}}) + E_{corr}(q),
\end{eqnarray}
where $E_{\text{tot}}[X^{q}]$ is the total energy of the supercell containing the defect $X$ in charge state $q$; $E_{\text{tot}}[\text{hBN}]$ is the total energy of pristine hBN; The term $n_i$ represents the number of atoms added or removed from the system, with $\mu_i$ being a chemical potential of atom species $i$; The Fermi level $\epsilon_{\text{Fermi}}$ corresponds to the electronic chemical potential, referenced to the valence band maximum (VBM) of pristine hBN, denoted by $\epsilon_{\text{vbm}}$. The final term, $E_{corr}(q)$, is a correction for spurious electrostatic interactions arising from periodic boundary conditions in charged supercells \cite{10.1103/PhysRevLett.102.016402}. In this work, we employed the Spinney Python code for charge correction \cite{10.1016/j.cpc.2021.107946}. Although charge states $q$ greater than $+$3 or $q$ less than $-$3 may be possible for certain defects, they were not considered in this work. Future studies may extend the analysis to include these regimes in order to identify additional stable defects and potential quantum emitters.\\
\indent By comparing the formation energies across charge states, we identify the most stable charge configuration for each defect, which is then used in all subsequent ground- and excited-state property calculations.

\subsection{Excited-state properties}
\subsubsection{Excited-state electronic structures}
To simulate excited-state configurations, we employ the so-called $\Delta$SCF method, a constrained DFT approach in which the occupation of electronic states is manually controlled~\cite{10.1103/RevModPhys.61.689}. In this method, one electron is explicitly promoted to an excited state, allowing us to model the electron excitation. \\
\indent For spin-polarized systems, such as defects with doublet or triplet ground states, two types of spin-conserving excitations may be possible: spin-up to spin-up or spin-down to spin-down. Both transitions are considered in this work where relevant. In contrast, for singlet defects, only one transition channel is evaluated unless otherwise specified. This ensures consistency with spin-selection rules and captures the dominant optical transitions for each defect configuration.

\subsubsection{Zero phonon line (ZPL) calculation}
In experimental photoluminescence spectra, the sharpest and most prominent peak is typically associated with the ZPL, which corresponds to a purely electronic transition occurring without the involvement of phonons. In this work, the ZPL is calculated as the total energy difference between the fully relaxed excited-state and ground-state geometries as follows:
\begin{equation}
    \text{ZPL} = E_{total,excited} - E_{total,ground}.
\end{equation}
While the zero-point vibrational energies (ZPE) can slightly shift the ZPL, typically in the range of a few meV to 0.1 eV, depending on the defect \cite{10.1038/s41524-023-01135-z,10.1103/PhysRevB.99.165201}, they are neglected in this study under the assumption that ZPE contributions approximately cancel between the ground and excited states. To ensure spin conservation, we treat spin-up and spin-down transitions separately. For certain defects, this results in two distinct spin-conserving excitation pathways, each yielding its own ZPL.

\subsubsection{Configuration coordinates}
Due to the differences in electronic occupation between the ground and excited states, the atomic geometries of a defect can undergo significant structural relaxation upon excitation. This structural change can be quantified by the so-called configuration coordinate ($q_k$), which is defined as
\begin{equation}
    q_k = \sum_{\alpha,i \in {x,y,z}}\sqrt{m_\alpha}\left(R_{e,\alpha i} - R_{g,\alpha i}\right)\Delta r_{k,\alpha i},
\end{equation}
where $\alpha$ indexes the atoms; $i$ denotes the Cartesian components $(x, y, z)$; and $m_{\alpha}$ is the mass of atom $\alpha$. The terms $R_{g,\alpha i}$ and $R_{e,\alpha i}$ correspond to the equilibrium positions of atom $\alpha$ in the ground and excited states, respectively; and $\Delta r_{k,\alpha i}$ is the normalized eigenvector of the phonon mode $k$, indicating the direction of atomic displacement.

\subsubsection{HR and DW factors}
The Huang-Rhys (HR) factor ($S$) can be used to quantify the strength of the electron-phonon coupling. This can be computed from the partial HR factor $s_k$ of each phonon mode $k$, defined as
\begin{equation}
    s_k = \frac{\omega_k q_k^2}{2\hbar}, \label{eq:s_k}
\end{equation}
where $\omega_k$ is the frequency of the phonon mode $k$; and $q_k$ is the configuration coordinate displacement corresponding to that mode, as defined previously. Then the HR factor is obtained by summing over all phonon modes, and its spectral distribution can be expressed as
\begin{equation}
    S(\hbar\omega) = \sum_k s_k\delta(\hbar\omega - \hbar\omega_k), \label{eq:S}
\end{equation}
where $\delta$ is the Dirac delta function. In addition to $S$, the Debye–Waller (DW) factor is also computed to characterize the fraction of emission occurring in the ZPL, which is given by DW = $e^{-S}$.

\subsubsection{Absorption and photoluminescence spectra}
In this work, the photoluminescence $L(\hbar\omega)$ spectrum is calculated from the full-phonon simulation \cite{10.1088/1367-2630/16/7/073026}, which can be obtained from
\begin{equation}
    L(\hbar\omega) = C\omega^3 A(\hbar\omega),
\end{equation}
where $C$ is a normalization constant (typically determined by fitting to experimental data), and $A(\hbar\omega)$ is the optical spectral function that captures both ZPL and phonon sidebands, which is computed as
\begin{equation}
    A(E_{ZPL} - \hbar\omega) = \frac{1}{2\pi}\int_{-\infty}^{\infty}G(t)\exp(-i\omega t-\gamma|t|) dt, \label{eq:PL}
\end{equation}
where $\gamma$ is a broadening parameter, and $G(t)$ is the generating function defined as
\begin{equation}
    G(t) = \exp(S(t) - S(0)).
\end{equation}
Here, $S(t)$ is the time-dependent spectral function given by
\begin{equation}
    S(t) = \int_0^\infty S(\hbar\omega)\exp(-i\omega t)d(\hbar\omega),
\end{equation}
where $S(\hbar\omega)$ represents the phonon-resolved Huang–Rhys factor as described in the previous section. The convergence test of the supercell size is shown in Supplementary S1.

\subsubsection{Transition rate and lifetime}
A radiative transition rate quantifies how quickly a transition occurs between two defect states with the same spin polarization. It is given by
\begin{equation}
\Gamma_{\mathrm{R}}=\frac{n_D e^2}{3 \pi \epsilon_0 \hbar^4 c^3} E_0^3 \mu_{\mathrm{e}-\mathrm{h}}^2,
\end{equation}
where $\Gamma_{\mathrm{R}}$ is the radiative transition rate; $e$ is the elementary charge; $\epsilon_0$ is vacuum permittivity; $E_0$ denotes the ZPL energy; $\mu_{\mathrm{e}-\mathrm{h}}^2$ is the modulus square of the dipole moment computed by Eq.~\eqref{eq:dipole}; and $n_{\mathrm{D}}$ is 1.85, which is the refractive index of the host hBN in the visible \cite{10.1039/c9nr04269e}. \\
\indent Since the refractive index of hBN can vary between samples or experimental conditions, our database provides an interactive feature that allows users to input a custom refractive index. The transition rate and corresponding radiative lifetime are automatically updated based on the user-defined value. The radiative lifetime ($\tau_R$) is computed as the inverse of the transition rate
\begin{equation}
    \tau_R = \frac{1}{\Gamma_{\mathrm{R}}}.
\end{equation}

\subsubsection{Transition dipole moment and polarization}
The polarization characteristics of excitation and emission processes can be derived from the transition dipole moments.~To compare theoretical predictions with polarization-resolved PL experiments, both excitation and emission dipoles are rotated by 90$^\circ$ and mapped modulo 60$^\circ$ to obtain the nearest alignment relative to the hexagonal crystal axes. Although the hexagonal symmetry is 120$^\circ$, angles beyond 180$^\circ$ are equivalent due to rotational symmetry. To ensure consistency with experimental conditions, dipole vectors are projected onto the $xy$-plane of the hBN crystal, allowing the computation of the in-plane polarization visibility.\\
\indent The transition dipole moments are calculated from the electronic wavefunctions obtained after structural relaxation. These wavefunctions are extracted using the Python package PyVaspwfc~\cite{pyWave}. However, since the wavefunctions in the ground and excited states originate from different geometries, two distinct dipole types must be considered: excitation and emission dipoles. The excitation dipole corresponds to the transition from the wavefunction of the optimized ground state to that of the optimized excited state. In contrast, the emission dipole corresponds to the transition from the optimized excited-state wavefunction to the optimized ground-state wavefunction. The transition dipole moment can be obtained from
\begin{equation}
    \boldsymbol{\mu} = \frac{i\hbar}{(E_{f} - E_{i})m}\bra{\psi_{f}}\textbf{p}\ket{\psi_{i}},
\label{eq:dipole}
\end{equation}
where $\psi_{i/f}$ are the initial and final wavefunctions; $E_{i/f}$ are the corresponding eigenvalues; $m$ is the electron mass; and $\mathbf{p}$ is the momentum operator. Because the involved wavefunctions come from geometrically distinct configurations, a modified version of PyVaspwfc~\cite{10.1088/1361-648X/ab94f4} is used to correctly evaluate the dipole moments. The resulting dipole vector $\boldsymbol{\mu}$ may have components in all three Cartesian directions and can be expressed as
\begin{equation}
    \boldsymbol{\mu} = \abs{\mu_x}\hat{x} + \abs{\mu_y}\hat{y} + \abs{\mu_z}\hat{z}.
\end{equation}
If $\mu_z$ = 0, the transition dipole lies entirely in the $xy$-plane, indicating a purely in-plane polarization. This behavior is further analyzed using the linear in-plane polarization visibility, which quantifies the extent to which the emission or absorption is confined within the plane of the hBN crystal.

\subsection{Database repository}
The hBN defect database is stored as a single SQLite file, \texttt{hbn\_defects\_database.db}, in the GitHub repository \texttt{QCS-Theory/hBN-database}.  It contains a primary table, \texttt{updated\_data}, with the following schematics as listed in Tab.~\ref{tab:db-schema}.
\begin{table*}[t]
  \centering
  \begin{tabular}{p{7.5cm} p{7.5cm} l} 
    \toprule
    \textbf{Column name} & \textbf{Type and description} & \textbf{\texttt{Option}} \\
    \midrule
    Host & \texttt{TEXT}: host material identifier & -\\
    Defect & \texttt{TEXT}: defect chemical formula/code & -\\
    Defect name & \texttt{TEXT}: descriptive name of the defect & -\\
    Charge state & \texttt{INTEGER}: integer charge state & -\\
    Spin multiplicity & \texttt{TEXT}: spin configuration of defects & -\\
    Optical spin transition & \texttt{TEXT}: allowed optical spin transition & -\\
    Excitation properties: dipole\_x (Debye) & \texttt{REAL}: x-component of excitation dipole moment & \texttt{abs\_dipole\_x} \\
    Excitation properties: dipole\_y (Debye) & \texttt{REAL}: y-component of excitation dipole moment & \texttt{abs\_dipole\_y} \\
    Excitation properties: dipole\_z (Debye) & \texttt{REAL}: z-component of excitation dipole moment & \texttt{abs\_dipole\_z} \\
    Excitation properties: linear In-plane Polarization Visibility & \texttt{REAL}: Visibility & \texttt{abs\_visibility} \\
    Excitation properties: Intensity (Debye) & \texttt{REAL}: Strength of excitation transition dipole moment & \texttt{abs\_tdm} \\
    Excitation properties: Characteristic time (ns) & \texttt{REAL}: Time of excitation transition & \texttt{abs\_lifetime} \\
    Excitation properties: Angle of excitation wrt the crystal axis & \texttt{REAL}: excitation polarization angle & \texttt{abs\_angle} \\
    Emission properties: dipole\_x (Debye) & \texttt{REAL}: x-component of emission dipole moment & \texttt{ems\_dipole\_x} \\
    Emission properties: dipole\_y (Debye) & \texttt{REAL}: y-component of emission dipole moment & \texttt{ems\_dipole\_y} \\
    Emission properties: dipole\_z (Debye) & \texttt{REAL}: z-component of emission dipole moment & \texttt{ems\_dipole\_z} \\
    Emission properties: linear In-plane Polarization Visibility & \texttt{REAL}: Visibility & \texttt{ems\_visibility} \\
    Emission properties: Intensity (Debye) & \texttt{REAL}: Strength of emission transition dipole moment & \texttt{ems\_tdm} \\
    Emission properties: ZPL (eV) & \texttt{REAL}: ZPL energy & \texttt{ZPL} \\
    Emission properties: ZPL (nm) & \texttt{REAL}: ZPL wavelength & \texttt{ZPL\_nm} \\
    Emission properties: lifetime (ns) & \texttt{REAL}: Radiative lifetime of emission & \texttt{lifetime} \\
    Emission properties: Angle of emission wrt the crystal axis & \texttt{REAL}: emission polarization angle & \texttt{ems\_angle} \\
    Emission properties: Polarization misalignment (degree) & \texttt{REAL}: dipole misalignment angle & \texttt{misalignment} \\
    Emission properties: Configuration coordinate (amu$^{1/2}$/\AA) & \texttt{REAL}: $Q$ value & \texttt{Q} \\
    Emission properties: HR factor & \texttt{REAL}: $S$ value & \texttt{HR} \\
    Emission properties: DW factor & \texttt{REAL}: DW value & \texttt{DW} \\
    Emission properties: Ground-state total energy (eV) & \texttt{REAL}: total energy & \texttt{E\_ground} \\
    Emission properties: Excited-state total energy (eV) & \texttt{REAL}: total energy & \texttt{E\_excited} \\
    Ground-state structure & \texttt{BLOB}: atomic structure file (CIF file) & \texttt{structure\_ground} \\
    Excited-state structure & \texttt{BLOB}: atomic structure file (CIF file) & \texttt{structure\_excited} \\
    Ground-state electronic structure & \texttt{BLOB}: raw OUTCAR file (VASP format) for electronic structure & \texttt{band\_ground} \\
    Excited-state electronic structure & \texttt{BLOB}: raw OUTCAR file (VASP format) for electronic structure & \texttt{band\_excited} \\
    PL lineshape & \texttt{BLOB}: raw PL lineshape file with the broadening parameter ($\gamma$) in Eq.~\eqref{eq:PL} equal to 1 & \texttt{PL} \\
    Raman spectrum & \texttt{BLOB}: raw Raman spectrum file & \texttt{raman}  \\
    \bottomrule
  \end{tabular}
  \caption{Structure of the \texttt{updated\_data} table in \texttt{hbn\_defects\_database.db}. The leftmost column provides the database field names, and the central column specifies the corresponding field’s data type along with a concise description. The rightmost column lists the retrieval \texttt{option} keys used by \texttt{get\_database(option=["..."])}. The first six rows are always returned in the \texttt{get\_database()} function. See Sec.~\ref{sec:instructions} for instructions in detail.}
  \label{tab:db-schema}
\end{table*}

To provide access, we use the GitHub REST API \texttt{contents} endpoint.  Users type
\begin{verbatim}
GET https://raw.githubusercontent.com/
    QCS-Theory/hBN-database/main/
    hbn_defects_structure.db
\end{verbatim}
This returns the SQLite binary, which users can save locally and open with any SQLite-compatible tool.  Alternatively, in Python, one can call \texttt{get\_database(...)} function to load and (optionally) filter the data. See Sec.~\ref{sec:instructions} for step-by-step procedures.

\section{Results and Discussion}
In this section, we first analyze correlations among key photophysical parameters. We then investigate the impact of vacancy on PSB broadening. Finally, we present the overall distribution of hBN defects in a histogram.

\subsection{Correlation matrix}
\begin{figure*}
    \centering
    \includegraphics[width=1\linewidth]{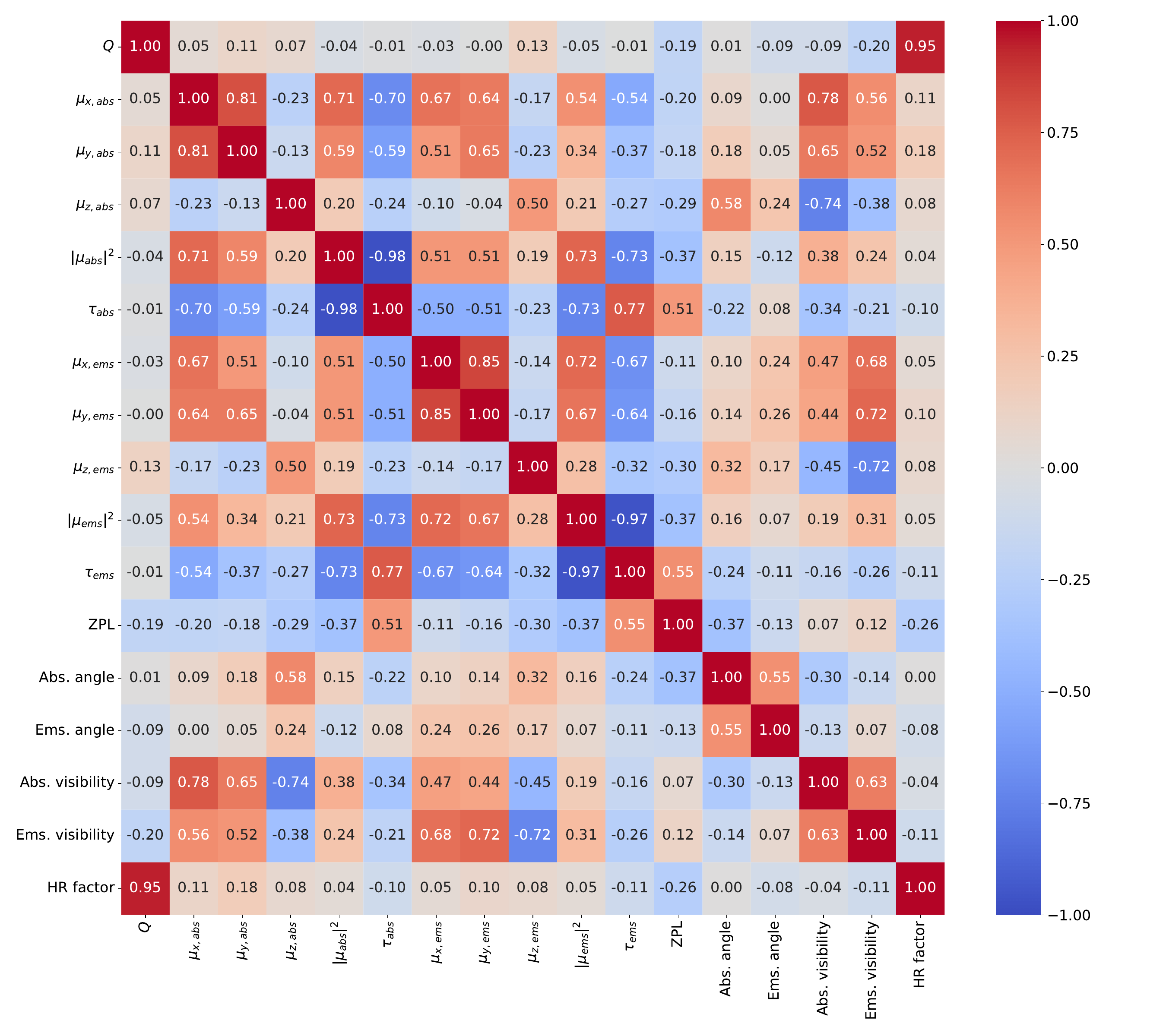}
    \caption{Correlation matrix based on the Spearman rank correlation method among 17 properties. High absolute values indicate strong correlations, while values close to zero imply weak or no correlation. Positive values represent increasing relationships between property pairs, whereas negative values indicate inverse relationships.}
    \label{fig:cor_matrix}
\end{figure*}

The correlation matrix among various photophysical parameters is computed using the Spearman rank correlation method, as shown in Fig.~\ref{fig:cor_matrix}. Overall, most parameters exhibit low correlation coefficients, indicating that they are largely independent of one another. However, notable correlations are observed between the configuration coordinate and the HR factor.\\
\indent These dependencies are expected and can be explained analytically. As shown in Eqs.~\eqref{eq:s_k} and \eqref{eq:S}, the HR factor is proportional to the square of the configuration coordinate, making their correlation inherently strong. This relationship validates the internal consistency of the computed dataset.

\subsection{Relationship between the number of vacancies and PSB broadening}
\begin{figure*}
    \centering
    \includegraphics[width=1\linewidth]{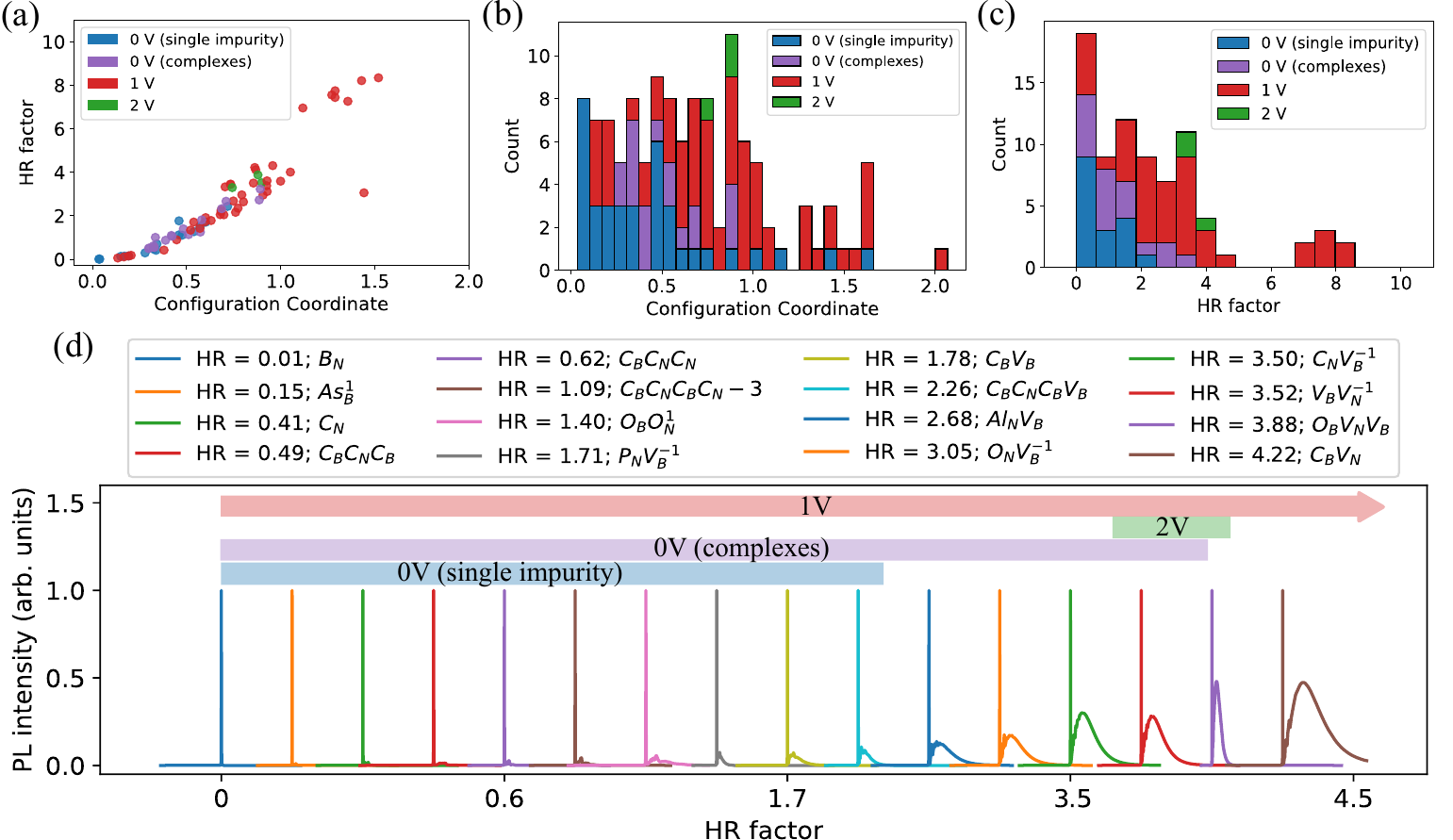}
    \caption{(a) Correlation between configuration coordinate and HR factor. (b) and (c) show histograms of the configuration coordinate between 0 and 2.5 amu$^{1/2}$\AA~and HR factor between 0 and 10, respectively. Defect types are color-coded as follows: blue for single-impurity defects without vacancies; purple for vacancy-free defect complexes (comprising multiple impurities); red for defects with a single vacancy; and green for defects with two vacancies. (d) displays selected but representative PL lineshapes categorized by the number of vacancies in each defect. The color stripe indicates the corresponding HR factor range for each defect category.}
    \label{fig:compare_HR_PSB}
\end{figure*}

We now investigate the behavior of the HR factor, which is directly related to PSB broadening and thus reflects the strength of electron–phonon coupling. In Fig.~\ref{fig:compare_HR_PSB}, defects are categorized into four groups based on their vacancy content: no vacancy with a single impurity, no vacancy with complexes, one vacancy, and two vacancies.\\
\indent Fig.~\ref{fig:compare_HR_PSB}(a) shows the correlation between the $Q$ and the HR factor. As expected from theory, the HR factor increases approximately with the square of $Q$, confirming the analytical relationship. Notably, defects without vacancies tend to exhibit lower values of both $Q$ and the HR factor. In contrast, defects with one or two vacancies span a wider range of values, with a significant fraction showing strong electron–phonon coupling (i.e., higher $Q$ and HR values). This trend is further supported by the histograms shown in Figs.~\ref{fig:compare_HR_PSB}(b) and ~\ref{fig:compare_HR_PSB}(c), where the distributions of the high HR factor and high $Q$ mostly belong to the defects with vacancies. \\
\indent PL lineshape comparisons in Fig.~\ref{fig:compare_HR_PSB}(d) further reinforce this observation. Defects without vacancies, including isolated point defects and complexes, typically exhibit the HR factors in the range of 0-2 and 0-3.8, respectively. In contrast, defects containing vacancies show significantly broader HR distributions: 0–12 for single-vacancy defects and 3.5–4 for those with two vacancies. As reflected in the PL lineshapes, vacancy-containing defects tend to display broader PSB. These results suggest that having more vacancies tends to induce stronger lattice distortion upon excitation compared to having no vacancy. This enhanced structural relaxation leads to greater geometric variation between the ground and excited states, thereby increasing the $Q$ and, consequently, the HR factor.

\subsection{Distribution of hBN defects}
\begin{figure*}
    \centering
    \includegraphics[width=1\linewidth]{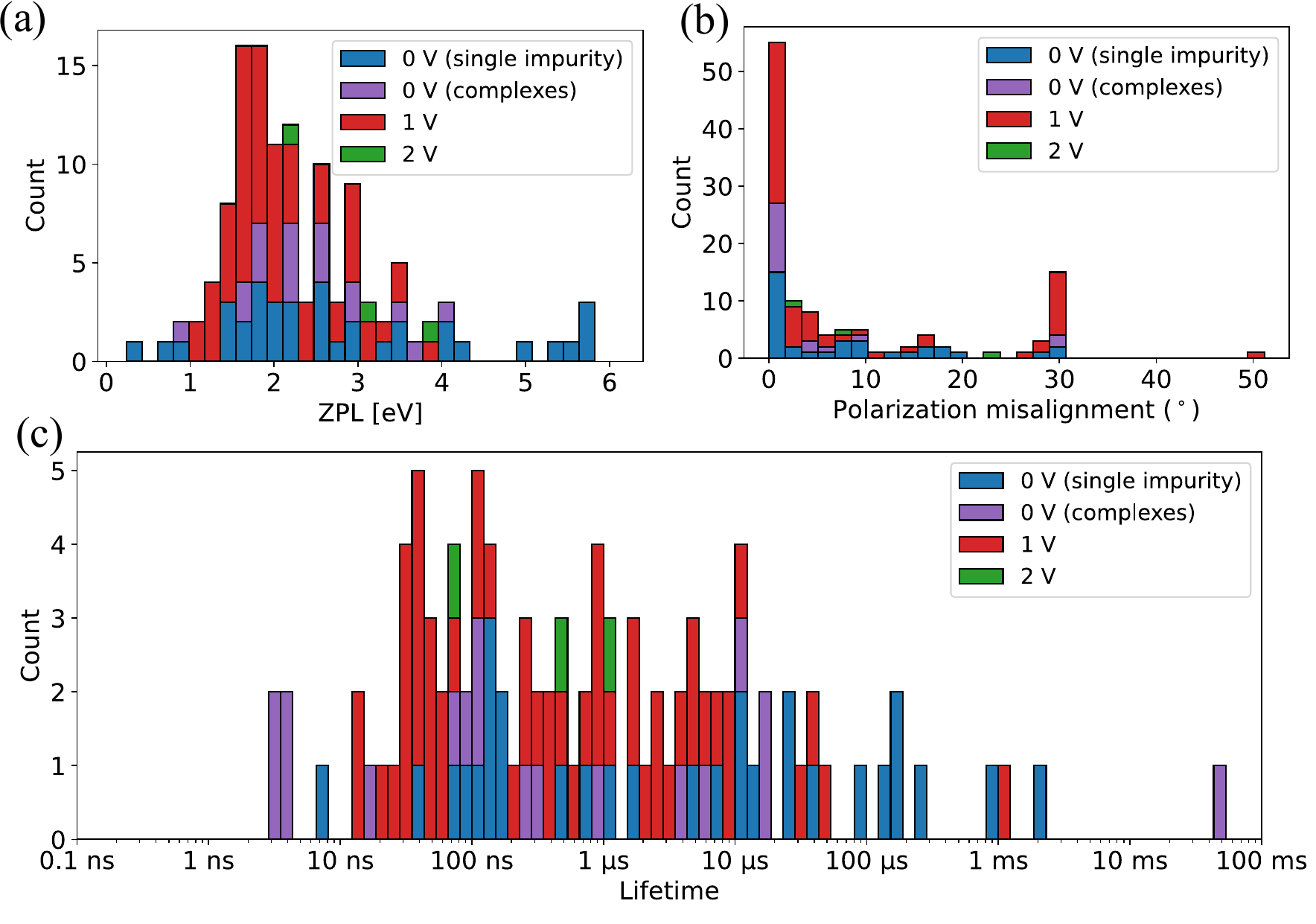}
    \caption{(a) Distribution of ZPL energies. (b) Distribution of polarization misalignment angles. (c) Distribution of radiative lifetimes. Blue bins (0V) represent single-impurity defects without vacancies; purple bins (0V) represent vacancy-free defect complexes (containing more than two impurities); red bins (1V) correspond to defects with one vacancy; and green bins (2V) represent defects containing two vacancies.}
    \label{fig:distribution}
\end{figure*}

To further analyze trends across different defect classes, Fig.~\ref{fig:distribution}(a) presents the distribution of ZPL energies for defects categorized by the number of vacancies. The results suggest that there is no clear correlation between ZPL energy and the number of vacancies: all three classes (no vacancy, one vacancy, two vacancies) span a wide ZPL range. Interestingly, vacancy-free defects with a single impurity appear to populate both the low and high ends of the ZPL spectrum.\\
\indent At the lowest ZPL bin, the O\textsubscript{N} defect is found, with a ZPL energy of 0.24 eV. However, when taking defect formation energies into account, this configuration is energetically less favorable compared to its positively charged counterpart (O$_\text{N}^{+1}$), suggesting that such a low-energy transition may not be experimentally observable under typical conditions.\\
\indent On the opposite end, the Ga\textsubscript{B} defect exhibits a ZPL of 5.82 eV, which is the highest in the dataset. This configuration is thermodynamically stable based on its defect formation energy. However, its electronic structure does not feature well-defined two-level mid-gap states. This behavior can be attributed to the fact that Ga and B belong to the same group in the periodic table, resulting in no difference in the number of valence electrons. Consequently, the electronic transition involved in Ga\textsubscript{B} closely resembles a near-band-edge transition, yielding a ZPL value approaching the intrinsic bandgap of hBN.\\
\indent For defects containing complexes with one or more vacancies, the ZPL energies consistently fall within the 1–4 eV range. This trend can be attributed to the nature of vacancies, which typically introduce localized electronic states within the mid-gap region of hBN. Given the intrinsic bandgap of approximately 6 eV, these vacancy-induced states often lie near the center of the gap. As a result, optical transitions between such defect levels commonly span energy differences of 1–4 eV, leading to ZPL values confined within this range.\\
\indent Fig.~\ref{fig:distribution}(b) presents the distribution of misalignment between excitation and emission polarizations. The results show that there is no correlation between the misalignment and the number of vacancies. For the majority of defects, excitation and emission dipoles are nearly parallel (minimal misalignment). Since the polarization angles are mapped modulo 60$^\circ$ to account for the hexagonal crystal symmetry, most values cluster near 0$^\circ$. However, one obvious outlier with a misalignment of 51$^\circ$ is identified as the O$_\text{N}$V$_\text{B}^{-1}$ defect. Further analysis reveals that this large offset arises from a strong difference in polarization visibility: the excitation dipole is primarily in-plane and linearly polarized, while the emission dipole has an out-of-plane component. As such, the polarization misalignment is substantial. \\
\indent Lastly,~turning to the radiative lifetime distribution shown in Fig.~\ref{fig:distribution}(c), the result suggests that most hBN defects exhibit lifetimes ranging from approximately 1 ns to 100 $\mu$s. Furthermore, there is no clear correlation between the lifetime and the number of vacancies. More broadly, no systematic relationship is observed between the number of vacancies and other key defect properties, as further illustrated in Supplementary S2. Full example calculations of defect properties in the database are given in Supplementary S3.

\section{Programmatic Retrieval of the hBN Defect Database via the GitHub API}
\label{sec:instructions}
In this section, we describe a procedure for accessing the hBN defect database directly from our GitHub repository using the GitHub REST API, apart from accessing our website \url{https://h-bn.info}. We provide a Python module, \texttt{get\_hBN\_defects\_database.py}, which automates (i) accessing the raw SQLite database and (ii) extracting a user‐specified subset of entries.~This functionality enables seamless retrieval of the complete dataset, thereby facilitating integration with machine learning workflows. The step-by-step procedure is as follows:
\begin{enumerate}
  \item \textbf{Obtain the retrieval script.} \\
    Download \texttt{get\_hBN\_defects\_database.py} and place it in your working directory.
  
  \item \textbf{Import the retrieval function.} \\
    In your Python script or interactive session, import the function
    \begin{verbatim}
from get_hBN_defects_database import get_database
    \end{verbatim}

  \item \textbf{Fetch and filter the database.} \\
    Invoke \texttt{get\_database()} with the desired filtering criteria\begin{verbatim}data = get_database(
    option=["ZPL"],
    host=["monolayer", "bulk"],
    spin_multiplicity=["singlet", "doublet",
    "triplet"],
    charge_state=[-2, -1, 0, 1, 2],
    optical_spin_transition=["up", "down"],
    value_range=(2.0, 4.0),
    download_db=False
)\end{verbatim}
    The keyword arguments perform the following functions:
    \begin{description}[leftmargin=1.5em]
      \item[\texttt{option}] 
        Specifies which database columns to return. The complete set of valid keys is listed in Tab.~\ref{tab:db-schema}. To retrieve all columns, use
        \begin{verbatim}
option = ["all"]
        \end{verbatim}
      \item[\texttt{host}] 
        selects between the monolayer and bulk hBN datasets. By default, both are returned.
      \item[\texttt{spin\_multiplicity}] 
        filters defects by their spin multiplicity. If omitted, all multiplicities are included.
      \item[\texttt{charge\_state}] 
        filters defects by charge state. Defaults to all if unspecified.
      \item[\texttt{optical\_spin\_transition}] 
        filters by optical spin transition (i.e.\ ``up" refers up to up or ``down" refers to down to down). Both are returned if this argument is not provided.
      \item[\texttt{value\_range}] 
        restricts the numeric range of the selected property. When omitted, no range filtering is applied.
      \item[\texttt{download\_db}] 
        If set to \texttt{True}, one can download the raw SQLite database file (named \texttt{hbn\_defects\_<options>.db}) to the working directory.
    \end{description}

  \item \textbf{Inspect the results.} \\
    The object \texttt{data} is a \texttt{pandas.DataFrame} containing the filtered subset, ready for downstream analysis or visualization. If \texttt{download\_db=True}, the corresponding \texttt{.db} file will also be saved locally.
\end{enumerate}
The complete code examples are available in Supplementary S4.

\section{Further Updates to the Database}
In addition to the completely new database of both ground and excited state properties in bulk hBN presented in this work, our hBN defects database has also undergone other updates since its initial release. This includes
\begin{enumerate}
    \item \textbf{Version 1.1}: Incorporation of configuration coordinate data for 277 monolayer hBN defects.
    \item \textbf{Version 1.2}: Inclusion of 554 total energy values for each structure, covering both ground- and excited-state configurations for monolayer hBN defects.
    \item \textbf{Version 1.3}: Inclusion of Raman spectra for 100 defects in hBN monolayers obtained from Ref.~\cite{arxiv:2502.21118}
\end{enumerate}

\section{Conclusion}
In summary, we have established an extended database of hBN defects that includes a comprehensive set of defects in bulk hBN, along with their excited-state photophysical properties. In addition to our previous work focusing on monolayer hBN and ground-state characteristics, this new database incorporates additional defect types in bulk hBN, charge-state analysis, and excited-state properties.~Key properties such as ZPL, PL lineshape, absorption lineshape, HR factor, interactive radiative lifetimes, transition dipole moments, and polarization characteristics have been systematically computed.\\
\indent From the current dataset, we computed a correlation matrix and found that only the configuration coordinate and HR factor are strongly correlated, consistent with their analytical relationship.~Further analysis showed that defects with vacancies tend to exhibit larger lattice distortions, leading to higher configuration coordinates and HR factors, resulting in broader PSBs. This suggests that broadened PSBs in low-temperature PL may serve as a possible indicator of vacancy-containing defects. Moreover, we analyzed the distribution of all parameters and found no additional strong correlations. \\
\indent We stress that the primary goal of our work is to construct a large database covering comprehensive properties of many different defects. This is to serve as a screening tool that helps rule out unlikely defects and narrow down promising candidates. This work does not take into account symmetry reduction effects such as the Jahn–Teller distortion; therefore, the specific values of the ZPL may be shifted. Such effects should be carefully considered when determining a defect as a promising candidate. For instance, to use our database, if an experiment reports a ZPL at 2.2 eV, one should focus on defects with calculated ZPLs between 2.0 and 2.4 eV, taking into account the finite accuracy of DFT calculations. This approach effectively eliminates many candidates and suggests the most plausible ones.\\
\indent Finally, all data are made publicly available through an online platform \url{https://h-bn.info}, alongside the GitHub REST API. This enables programmatic retrieval of the complete SQLite database or user-specified subsets of defect properties. The API returns raw or filtered \texttt{.db} files on demand, and can be seamlessly integrated into automated pipelines and ML workflows. This resource is therefore intended to assist researchers in experimental interpretation, accelerate defect identification, and facilitate data-driven discovery in the field of solid-state quantum emitters.  

\section*{Data availability}
All raw data from this work is available from the authors upon reasonable request. The database with browsing interfaces is freely accessible at \url{https://h-bn.info}. The SQLite database file can be downloaded via the GitHub REST API.

\section*{Notes}
The authors declare no competing financial interest. We acknowledge the assistance of OpenAI’s ChatGPT in managing code snippets. All final decisions on integration and adaptation of those suggestions were made by the authors.

\begin{acknowledgments}
This research is part of the Munich Quantum Valley, which is supported by the Bavarian state government with funds from the Hightech Agenda Bayern Plus. This work was funded by the Deutsche Forschungsgemeinschaft (DFG, German Research Foundation) under Germany's Excellence Strategy- EXC-2111-390814868 (MCQST) and as part of the CRC 1375 NOA project C2. The authors acknowledge support from the Federal Ministry of Research, Technology and Space (BMFTR) under grant number 13N16292 (ATOMIQS). S.S. acknowledges research funding by Mahidol University (Fundamental Fund FF-111/2568: fiscal year 2025 by the National Science Research and Innovation Fund (NSRF)). The authors gratefully acknowledge the Gauss Centre for Supercomputing e.V.\ (www.gauss-centre.eu) for funding this project by providing computing time on the GCS Supercomputer SuperMUC-NG at Leibniz Supercomputing Centre (www.lrz.de) and on its Linux-Cluster.
\end{acknowledgments}


\bibliography{main}
\end{document}